\documentclass{appolb}
\usepackage{epsfig}
\usepackage{amsmath}
\usepackage{amssymb}
\newcommand{\bc}{\begin{center}}
\def\ec{\end{center}}
\newcommand{\Pom}{\mbox{$I\!\!P$}}
\newcommand{\aP}{\mbox{$\alpha_{\Pom}$}}   

\begin{document}
\title{The Rise of the Proton Structure Function $F_2$ \\ 
Towards Low $x$
\thanks{Presented at DIS2002, Krak\'ow, 30.4. - 4.5. 2002}%
}
\author{J\"org Gayler, DESY
\address{on behalf of the H1 collaboration}
}
\maketitle
\begin{abstract}
Results on the derivative of $\log (F_2)$ with respect to $\log (x)$ at
 fixed $Q^2$ are presented. The measured derivatives are within errors
 independent of $x$ for $Q^2 \geqq 0.85$ GeV$^2$
 and increase linearly with
 $\log (Q^2)$ for $10^{-4} \leqq x \leqq 0.01$ and $Q^2 \gtrapprox 3$ GeV$^2$.
 The results are based on preliminary and published H1 data which
 at $Q^2$ below 2 GeV$^2$ are combined
 with NMC and ZEUS data.
\end{abstract}
  
\section{Introduction}
 The rise of the proton structure function $F_2$ towards small Bjorken $x$
  has been discussed since the existence of QCD. In the double asymptotic limit
  (large energies, i.e. small $x$, and large photon virtualities $Q^2$)
  the DGLAP evolution equations ~\cite{DGLAP} can be solved
~\cite{DeRujula:rf}
   and $F_2$ is
   expected to rise approximately like a power of $x$ towards low $x$.  
  A power like behaviour is also expected in the BFKL approach ~\cite{bfkl}.
  However, it soon was discussed ~\cite{Gribov:1981ac}
  that this rise may eventually be limited
  by gluon self interactions in the nucleon, or more generally due to 
  unitarity constraints.

  Experimentally this rise towards small $x$ was first observed
  in 1993 in the HERA data ~\cite{Abt:1993cb}.
  Meanwhile the precision of the $F_2$ data is much improved and the rise
  can be studied in great detail.
   
\section{Procedure}
 
  The low $x$ behaviour of $F_2$ at fixed $Q^2$ is studied locally
  by the measurement of the derivative 
  $\lambda \equiv -(\partial \ln F_2/\partial \ln x)_{Q^2}$ as function of
  $x$ and $Q^2$. 
  The results 
  are based on preliminary H1 $F_2$ data
  presented to this conference ~\cite{lastovicka} covering the range
  $0.5 < Q^2 < 3.5$ GeV$^2$ and  
 published H1 data ~\cite{Adloff:2000qk},~\cite{Adloff:2001rw}
  which cover the range  $1.5 < Q^2 < 150$ GeV$^2$. 
  The low $Q^2$ $F_2$ data were obtained by shifting the $ep$
  interaction vertex by 70 cm in proton beam direction ~\cite{lastovicka}.
  At $Q^2 < 2$ GeV$^2$ the H1 data are also shown combined with data of
 NMC ~\cite{nmc}
 and ZEUS ~\cite{Breitweg:2000yn}. 
 The derivative
 $\lambda(x,Q^2)$ is evaluated using data points at adjacent values of $x$
 at fixed $Q^2$ taking into account error correlations and $x$ spacing
 corrections. The derivatives are compared with the next to leading order
 (NLO) QCD fit to the H1 cross section data ~\cite{Adloff:2000qk} and a
 ``fractal'' fit ~\cite{Lastovicka:2002hw}
 where self-similar properties of the proton structure are assumed.    

\section{Results}

\begin{figure}[ht]
 \begin{center}
\begin{picture}(200,336)
   \put(-15,220.){\epsfig{file=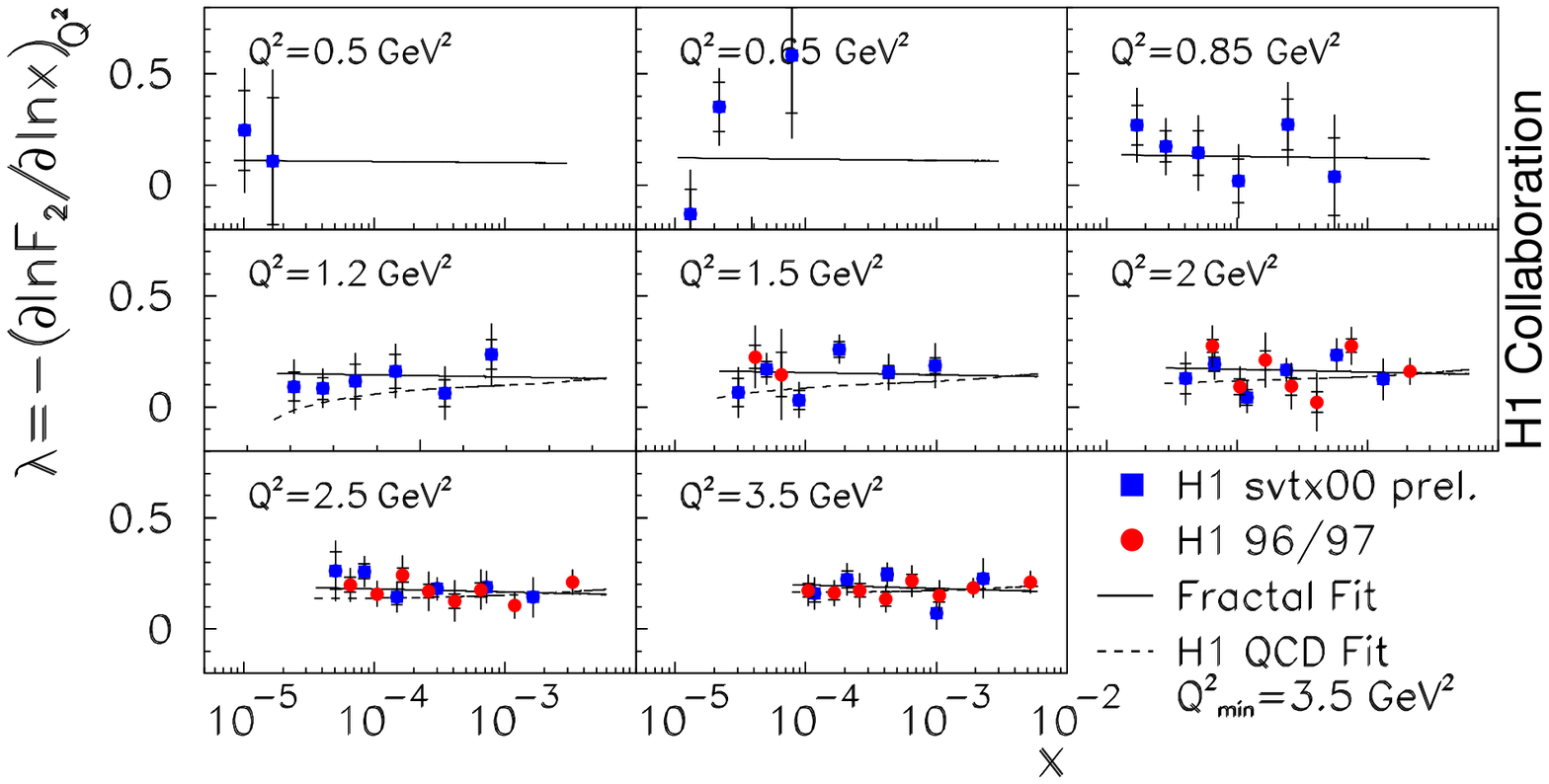,width=220pt}}
   \put(-15,5.){\epsfig{file=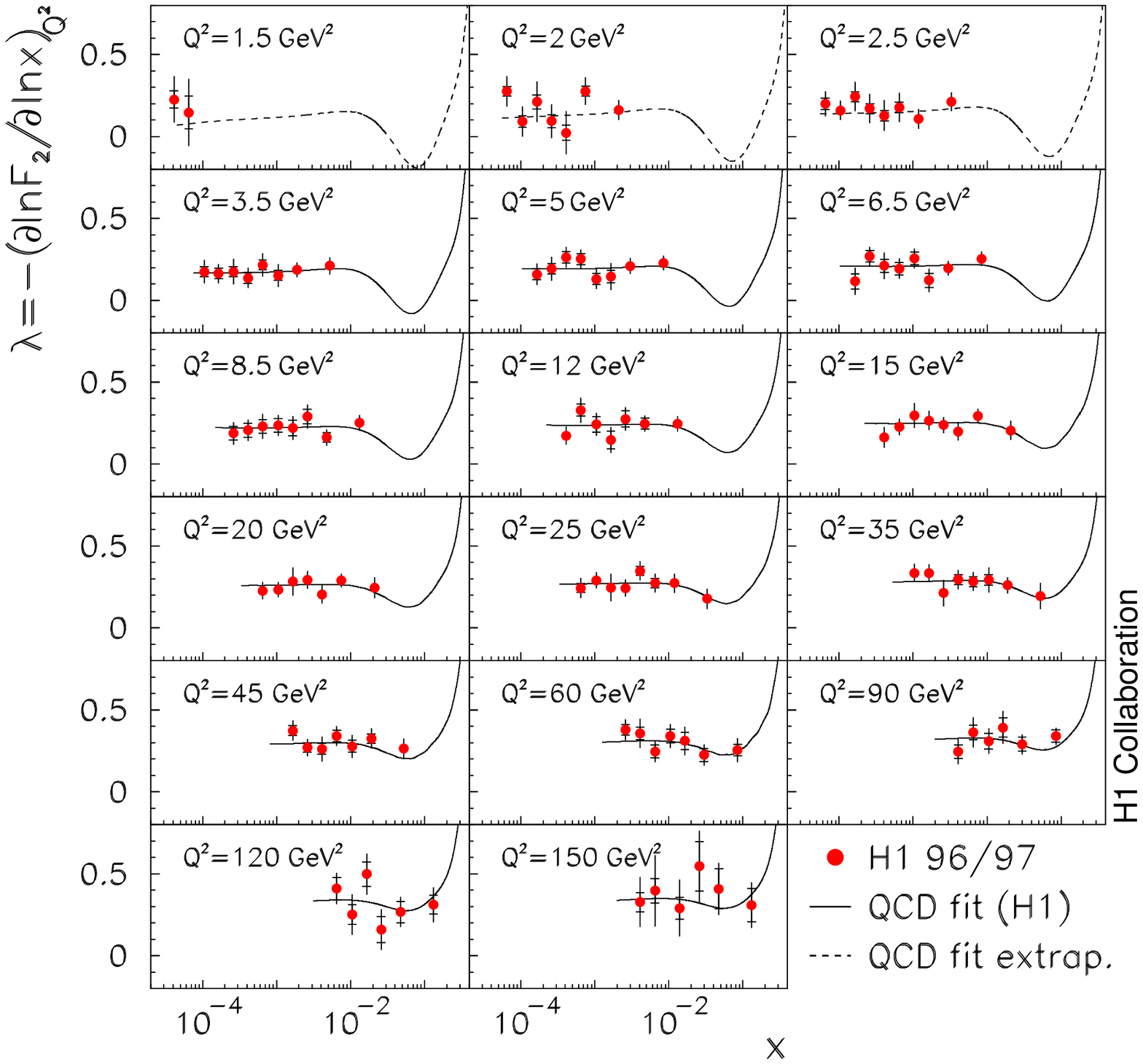,width=220pt}}
 \end{picture}
\caption{
 Derivative $\lambda=-(\partial \ln F_2/\partial \ln x)_{Q^2}$
compared with the QCD analysis
 of ref.~\cite{Adloff:2000qk} and a ``fractal'' fit ~\cite{Lastovicka:2002hw}
for $0.5 < Q^{2} < 3.5$ GeV$^2$ (upper plot)
 and for  $1.5 < Q^2 < 150$ GeV$^2$ (lower plot)  
\label{fig:der1}}
 \end{center}
\end{figure}

The $x$ and $Q^2$ dependence of 
  $\lambda=-(\partial \ln F_2/\partial \ln x)_{Q^2}$  
 is shown in Fig.~\ref{fig:der1}. 

\begin{figure}[ht]
 \begin{center}
\begin{picture}(200,315)
   \put(-25,105.){\epsfig{file=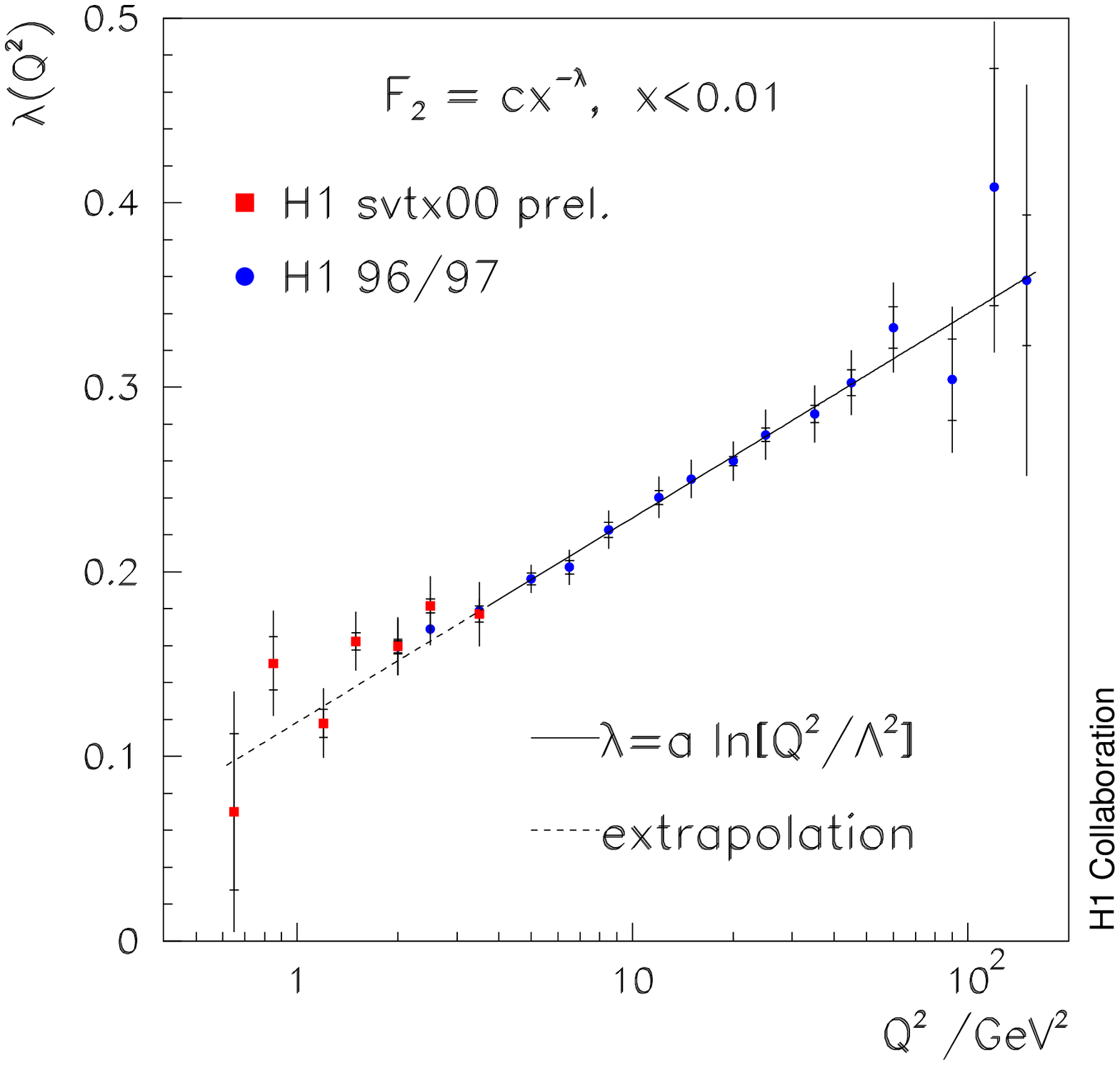,bbllx=21pt,bblly=155pt,bburx=546pt,bbury=655pt,angle=0,width=220pt,clip=}}
   \put(-25,0.){\epsfig{file=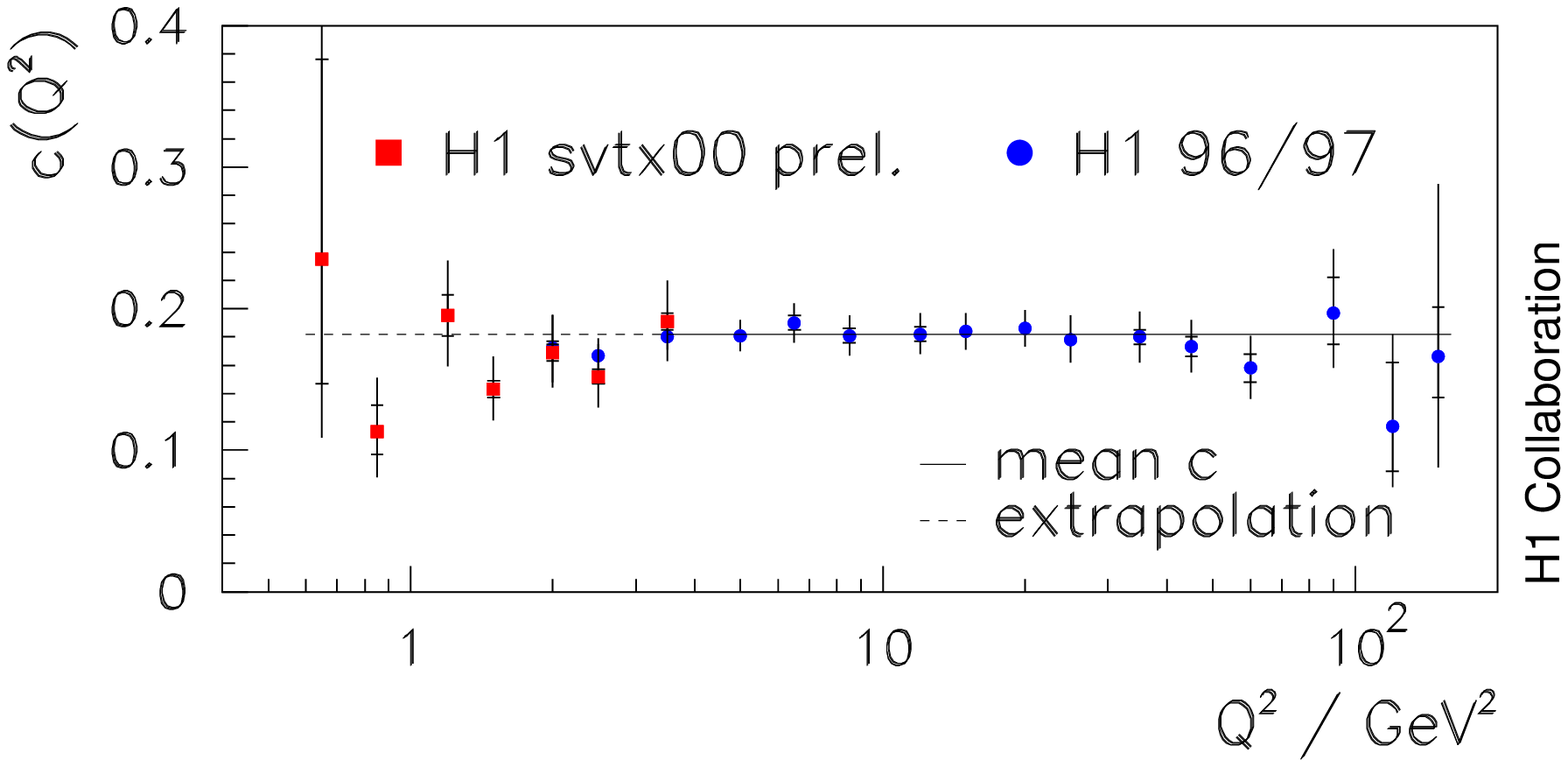,width=220pt}}
 \end{picture}
\caption{
$\lambda(Q^2)$ and $c(Q^2)$ from fits of the form
$F_{2} = c(Q^2) \cdot x^{-\lambda(Q^{2})}$
 to the H1 structure function data  
 ~\cite{Adloff:2000qk} and ~\cite{Lastovicka:2002hw}.
\label{fig:lamh1}}
 \end{center}
\end{figure}

 The new shifted vertex and the published data agree well
 in the overlap region.
  The derivative $\lambda$ is constant
  within
  experimental uncertainties for fixed $Q^{2}$ in the range $x < 0.01$,
  implying that the data are consistent with the power behaviour
  $F_{2} = c(Q^2) \cdot x^{-\lambda(Q^{2})}$.
  Fitting this form for each $Q^2$ bin to the data at $x < 0.01$,
  results in the $\lambda$ and $c$ values presented in
 Fig.~\ref{fig:lamh1}.

 The results show that the $F_{2}$ data at low $x$ for
 $Q^2 \gtrapprox 3.5$ GeV$^2$ can be well described by
 the very simple parameterisation
\begin{equation} \label{eq:f2p}
 F_2 = c \cdot x^{-\lambda(Q^2)}\;,\;\; {\rm with} \;\;
         \lambda(Q^2) = a \cdot \ln[Q^2/\Lambda^2]
\end{equation}
with $a = 0.0481 \pm .0013 \pm .0037$ and $\Lambda = 292 \pm 20 \pm 51$ MeV  
and  $c \approx 0.18$.

\begin{figure}[ht]
 \begin{center}
\begin{picture}(200,315)
   \put(-35,105.){\epsfig{file=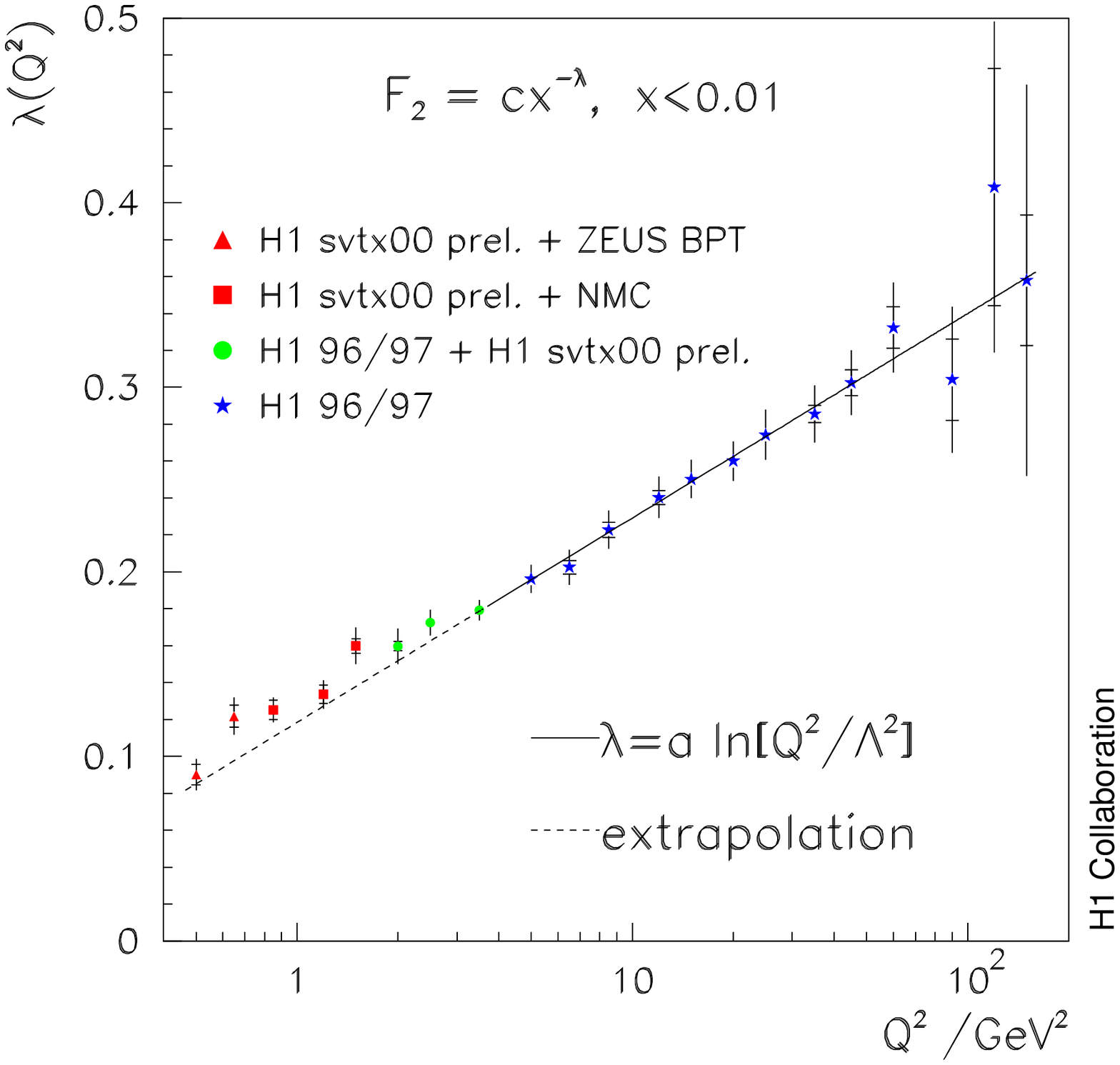,width=220pt}}
   \put(-35,0.){\epsfig{file=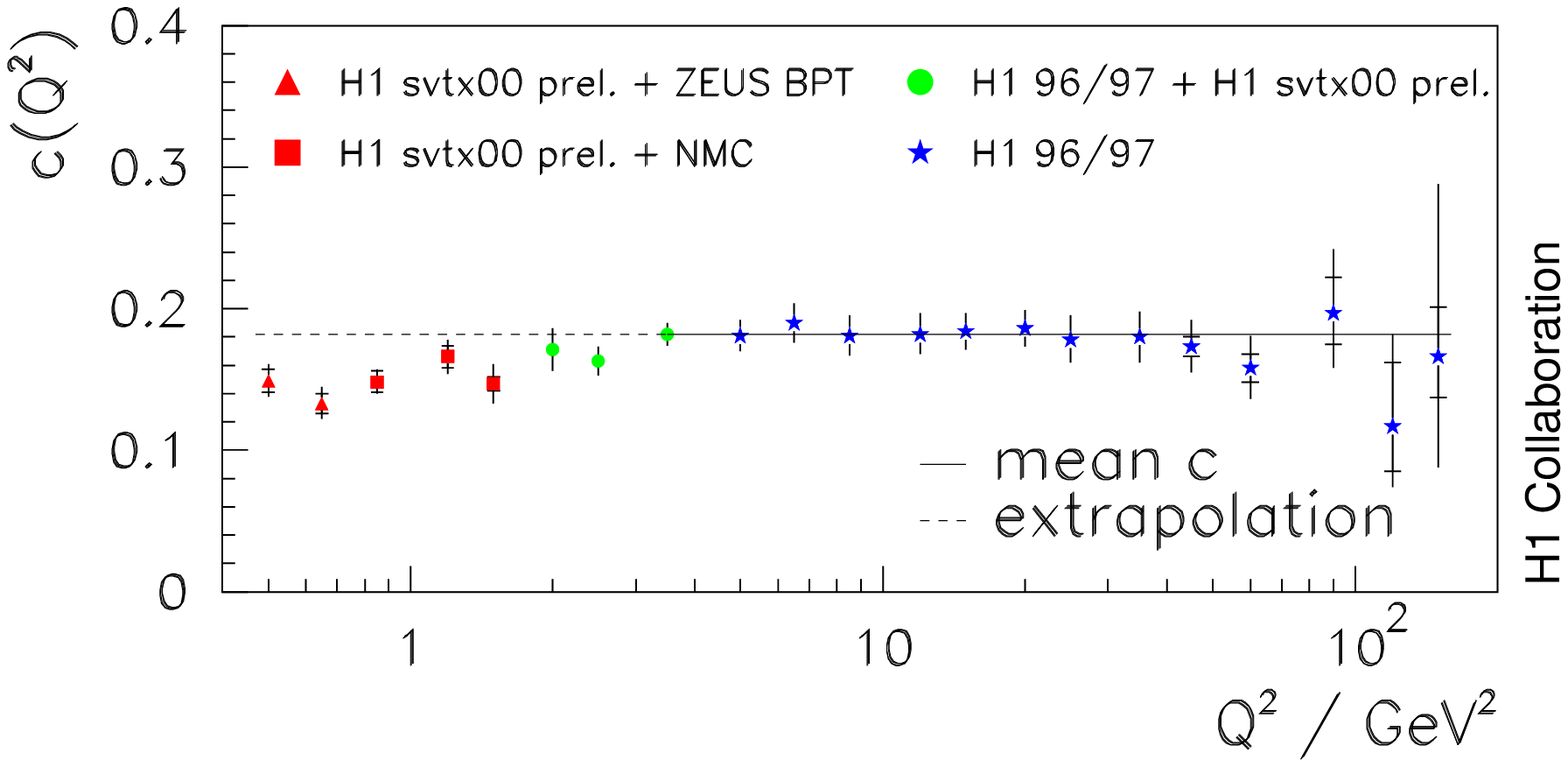,width=220pt}}
 \end{picture}
\caption{
$\lambda(Q^2)$ and $c(Q^2)$ from fits of the form
$F_{2} = c(Q^2) \cdot x^{-\lambda(Q^{2})}$
 combining the H1 structure function data of  
 ~\cite{Adloff:2000qk} and ~\cite{Lastovicka:2002hw} and the H1 data
 with data of NMC ~\cite{nmc} and ZEUS ~\cite{Breitweg:2000yn}. 
\label{fig:lammerge}}
 \end{center}
\end{figure}

 At low $Q^2$ the deviation of $\lambda$ from the logarithmic $Q^2$ dependence
 and the decrease of $c(Q^2)$ is more significant if the
 H1 data are combined with 
 NMC ~\cite{nmc}
 and ZEUS ~\cite{Breitweg:2000yn} data (see Fig.~\ref{fig:lammerge}).
                             
 The deviations from a simple constant respectively logarithmic behaviour
 occur at about such $Q^2$ values below which perturbative QCD fits
  (e.g.~\cite{Adloff:2000qk}) 
 are not supposed to be valid.
 At small $Q^2$ the structure function $F_2$ can be related
 to the total virtual
 photon absorption cross section by
\begin{equation} 
\sigma_{tot}^{\gamma^*p} = 4 \pi \alpha^2 F_2 /Q^2 \; \sim x^{-\lambda}/Q^2 
\end{equation} 
 where the total $\gamma^*p$ energy squared is given by $s = Q^2/x$. 
 For $Q^2 \rightarrow 0$ we can expect
   $c(Q^2) \rightarrow 0$ and
   $\lambda(Q^2) \rightarrow \approx 0.08$. The latter value corresponds 
   to the energy dependence of soft hadronic interactions
   $\sigma_{tot} \sim s^{\aP(0)-1}$ with $\aP(0)-1 \approx 0.08$
~\cite{Donnachie:1992ny}
   which is approximately reached at $Q^2 = 0.5$ GeV$^2$.

\section{Conclusion}
No significant deviation from the power behaviour
    $F_{2} \sim x^{-\lambda}$ at fixed $Q^2$ 
 is visible at
present energies and $Q^{2} \gtrsim 0.85$ GeV$^{2}$. More specifically:

\begin{itemize}
\item
 For $x < 0.01$ the derivative
 $\lambda \equiv -(\partial \ln F_2/\partial \ln x)_{Q^2}$ is independent
 of $x$ within errors.
\item
$\lambda$ is proportional to $\ln(Q^2)$ for $Q^2 \gtrsim 3$ GeV$^2$,
 i.e. in the pQCD region.
\item
Here the data can be very simply parametrised by 
$F_2 = c x^{-\lambda(Q^2)}$.
\item
At  $Q^2 \lesssim 3$ GeV$^2$ deviations from the logarithmic $Q^2$
dependence  of
$\lambda$ are observed.
\item
At low $Q^2 \;\; (Q^2 \lesssim 1$ GeV$^2)$ the energy rise is similar
 as in soft hadronic interactions.
\end{itemize}


\vspace*{-12.6pt}

\begin{thebibliography}{99}
\bibitem{DGLAP}
Y.~L.~Dokshitzer,
(In Russian),''
Sov.\ Phys.\ JETP {\bf 46} (1977) 641
[Zh.\ Eksp.\ Teor.\ Fiz.\  {\bf 73} (1977) 1216];     \\
V.~N.~Gribov and L.~N.~Lipatov,
Yad.\ Fiz.\  {\bf 15} (1972) 1218
[Sov.\ J.\ Nucl.\ Phys.\  {\bf 15} (1972) 675];   
Yad.\ Fiz.\  {\bf 15} (1972) 781
[Sov.\ J.\ Nucl.\ Phys.\  {\bf 15} (1972) 438]; \\
G.~Altarelli and G.~Parisi,
Nucl.\ Phys.\ B {\bf 126} (1977) 298.

\bibitem{DeRujula:rf}
A.~De Rujula et al,
Phys.\ Rev.\ D {\bf 10} (1974) 1649;  \\
R.~D.~Ball and S.~Forte,
Phys.\ Lett.\ B {\bf 335} (1994) 77.


\bibitem{bfkl}
E.~A.~Kuraev, L.~N.~Lipatov and V.~S.~Fadin,
Sov.\ Phys.\ JETP {\bf 44} (1976) 443
[Zh.\ Eksp.\ Teor.\ Fiz.\  {\bf 71} (1976) 840]; 
Sov.\ Phys.\ JETP {\bf 45} (1977) 199
[Zh.\ Eksp.\ Teor.\ Fiz.\  {\bf 72} (1977) 377];  \\
I.~I.~Balitsky and L.~N.~Lipatov,
Sov.\ J.\ Nucl.\ Phys.\  {\bf 28} (1978) 822
[Yad.\ Fiz.\  {\bf 28} (1978) 1597].

\bibitem{Gribov:1981ac}
L.~V.~Gribov, E.~M.~Levin and M.~G.~Ryskin,
Nucl.\ Phys.\ B {\bf 188} (1981) 555;
Phys.\ Rept.\  {\bf 100} (1983) 1; \\
A.~H.~Mueller and J.~w.~Qiu,
Nucl.\ Phys.\ B {\bf 268} (1986) 427.

\bibitem{Abt:1993cb}
I.~Abt {\it et al.}  [H1 Collaboration],
Nucl.\ Phys.\ B {\bf 407} (1993) 515; \\
M.~Derrick {\it et al.}  [ZEUS Collaboration],
Phys.\ Lett.\ B {\bf 316} (1993) 412.

\bibitem{lastovicka}
T.~Lastovicka, these proceedings.

\bibitem{Adloff:2000qk}
C.~Adloff {\it et al.}  [H1 Collaboration],
Eur.\ Phys.\ J.\ C {\bf 21} (2001) 33.

\bibitem{Adloff:2001rw}
C.~Adloff {\it et al.}  [H1 Collaboration],
Phys.\ Lett.\ B {\bf 520} (2001) 183.



\bibitem{nmc}
M.~Arneodo {\it et al.}  [New Muon Collaboration.],
Phys.\ Lett.\ B {\bf 364} (1995) 107;
Nucl.\ Phys.\ B {\bf 483} (1997) 3.

\bibitem{Breitweg:2000yn}
J.~Breitweg {\it et al.}  [ZEUS Collaboration],
Phys.\ Lett.\ B {\bf 487} (2000) 53.


\bibitem{Lastovicka:2002hw}
T.~Lastovicka,
arXiv:hep-ph/0203260 and these proceedings.

\bibitem{Donnachie:1992ny}
A.~Donnachie and P.~V.~Landshoff,
Phys.\ Lett.\ B {\bf 296} (1992) 227.
\end{thebibliography}
\end{document}